\documentclass[12pt]{article}
\usepackage{amsmath}
\usepackage{amssymb}
\input amssym.def
\input amssym

\newcommand{\be}{\begin{equation}}
\newcommand{\ee}{\end{equation}}

\title{Nariai--Bertotti--Robinson spacetimes as a building
material for one-way wormholes with horizons, but without
singularity}
\author{Nikolai V. Mitskievich\thanks{Physics Department, CUCEI,
University of Guadalajara, Guadalajara, Jalisco, Mexico.}
\thanks{Postal address: Apartado Postal 1-2011, C.P. 44100,
Guadalajara, Jalisco, M\'exico. E-mail:
mitskievich03@yahoo.com.mx}, Mar{\'\i}a G. Medina Guevara\\ and
H\'ector Vargas Rodr{\'\i}guez\thanks{C.U.Lagos de la UdeG,
Enr{\'\i}que D{\'\i}az de Le\'on S/N, Lagos de Moreno, Jal., C.P.
47460, M\'exico. E-mail: hv\!$_{{\rm -}\!}8$@yahoo.com}}
\date{~}
\begin{document}

\maketitle

\begin{abstract}
We discuss the problem of wormholes from the viewpoint of gluing
together two Reissner--Nordstr\"om-type universes while putting
between them a segment of the Nariai-type world (in both cases
there are also present electromagnetic fields as well as the
cosmological constant). Such a toy wormhole represents an example
of one-way topological communication free from causal paradoxes,
though involving a travel to next spacetime sheet since one has to
cross at least a pair of horizons through which the spacetimes'
junction occurs. We also consider the use of thin shells in these
constructions. Such a ``material'' for wormholes we choose taking
into account specific properties of the Nariai--Bertotti--Robinson
spacetimes.
\end{abstract}

In general relativity, the problem of wormholes is not more exotic
than that of black holes. In this talk we consider a simple toy
model which is still far from perfection which could however be
useful in better comprehension of the magnitude of the wormhole
problem.

The Nariai--Bertotti--Robinson (NBR) solution\cite{Bertotti,
Nariai1, Nariai2, Robinson, Exact2} (about the result of
Robinson\cite{Robinson} see however Ref. 3) can be described as
$ds^2= \hbox{\large e}^{2\alpha(r)}dt^2 -\hbox{\large
e}^{-2\alpha(r)}d r^2 -\lambda^2(d\vartheta^2+\sin^2\vartheta
d\varphi^2)$ where $\hbox{\large
e}^{2\alpha}=(k^2-\Lambda)r^2+Br+C$ and
$\lambda=\frac{1}{\sqrt{\Lambda+k^2}}$, $B$ and $C$ being
arbitrary constants, $\Lambda$ the cosmological constant, and $k$,
the (constant) electromagnetic field intensity. The
electromagnetic sources in Einstein's equations correspond to the
four-potential ${\cal A}=\sqrt{\frac{8\pi}{\kappa}}\left(akr\,dt+
\frac{kb} {(k^2+\Lambda)} \cos\vartheta\, d\varphi\right)$:
$T_{em}= \frac{k^2}{\kappa}\left( \theta^{(0)}\otimes\theta^{(0)}-
\theta^{(1)}\otimes\theta^{(1)}+\right.$\linebreak
$\left.\theta^{(2)}\otimes\theta^{(2)}
+\theta^{(3)}\otimes\theta^{(3)}\right)$ with $a=\sin\psi$,
$b=\cos\psi$, $\psi$ being an arbitrary constant, while (see a
general discussion in Ref. 4) $E=*(\theta^{(0)}\wedge*F)={\cal
A}_{0,r} \theta^{(1)},~~ B=*(\theta^{(0)}\wedge
F)=\frac{\Lambda+k^2}{\sin\vartheta} {\cal
A}_{3,\vartheta}\theta^{(1)}$ where $\theta^{(0)}=\hbox{\large
e}^{\alpha}dt,~~ \theta^{(1)}=\hbox{\large e}^{-\alpha}d r,~~
\theta^{(2)}=\lambda d\vartheta,~~
\theta^{(3)}=\lambda\sin\vartheta d\varphi$.

We consider pieces of NBR solutions with two horizons (null
compact hypersurfaces along whose generatrices $ds^2=0$, while $
\left| \frac{dt}{dr}\right| \rightarrow\infty $ on the horizons).
When $k^2>\Lambda>-k^2$, the two horizons are at $r=\pm r_0=\pm
1/\sqrt{k^2-\Lambda}$ with a non-stationary band between them and
static regions outside. Alternatively, when $k^2<\Lambda$, the
horizons are at $r=\pm r_0=\pm 1/ \sqrt{\Lambda-k^2}$ and
spacetime is static between them and non-stationary outside. We
now write these solutions in synchronous coordinates (see the
definition in the footnote on p. 62 of Ref. 4): \be \label{NBRsyn}
ds^2=dT^2-[{\cal E}^2-g_{00}(r(T,R))]dR^2-\frac{1}{\Lambda+k^2}[d
\vartheta^2 +\sin^2\vartheta d\varphi^2]. \ee Here ${\cal E}$ is
energy per unit mass of the geodesically moving test particle
identified with the observer. On horizons where $g_{00}=0$, no
singularities and degeneracy appear in the metric coefficients.
(This makes it unnecessary to apply the intrinsic prescription in
the Barrab\`es and Israel formalism. At the horizon there is then
used a thin null shell.\cite{Barrabes, Musgrave1, Musgrave2}) This
description also gives a unique junction of spacetimes and enables
the standard causal treatment of an infinite sequence of universes
in the Penrose diagram. In the case $k^2>\Lambda>-k^2$,
$g_{00}(r)=(k^2-\Lambda)r^2-1$; in the case $k^2<\Lambda$,
$g_{00}(r)=1-(\Lambda-k^2)r^2$.

As the outside worlds we consider the
Reissner--Nordstr\"om--Kottler (RNK) solutions\cite{Exact2} (those
of Reissner--Nordstr\"om, but with the cosmological term), \be
\label{RNKsyn} ds^2=dT^2-\left[{\cal E}^2-g_{00}(r)\right]dR^2-
r^2\left(d\vartheta^2 +\sin^2\vartheta d\varphi^2\right) \ee with
$g_{00}=1-\frac{2m_1}{r}+\frac{e^2_1}{r^2}- \frac{1}{3}\Lambda_1
r^2$ and $ g_{00}=1-\frac{2m_2}{r}+\frac{e^2_2}{r^2}
-\frac{1}{3}\Lambda_2 r^2$, $r=r(T,R)$. They are to be joined {\it
via} wormholes which belong to the NBR spacetimes, (\ref{NBRsyn}),
with $g_{00}(r)=\left(\frac{k^2-\Lambda}
{k^2+\Lambda}\right)\left[\left(\Lambda+k^2\right)r^2-1\right]$
(there is also $r(T,R)$, but with another dependence than in RNK),
when the cases $\Lambda>k^2$ and $k^2>\Lambda>-k^2$ are unified
via a scales change in $r$, so that the horizons correspond to
$r=\pm\lambda=\pm\frac{1}{\sqrt{\Lambda+k^2}}$ (the minus sign
does not spoil our considerations since it can be inverted when we
consider the junction of the NBR-wormhole with the `second' RNK
world at this horizon). At the horizons in synchronous coordinates
we put in (\ref{RNKsyn}) $g_{00}=0$ and substitute instead of $r$,
$\bar{r}=\bar{r}(m_1,e_1, \Lambda_1)= \bar{r}
(m_2,e_2,\Lambda_2)$, corresponding to anyone of the (three)
horizons of RNK, while in NBR (\ref{NBRsyn}) the only change at
the horizon is to put $g_{00}=0$. Hence we conclude that \be
[{\tilde g}_{\alpha\beta}]=0 ~~~ \Rightarrow ~~~
\frac{1}{\sqrt{k^2+\Lambda}}=\bar{r}\label{A1}. \ee The
electromagnetic stress-energy tensors read $ T_{\rm
NBR}=\frac{k^2}{\kappa}\left\{dT\otimes dT-\right.$\linebreak
$\left.{\cal E}^2dR\otimes dR
+\frac{1}{k^2+\Lambda}\left(d\vartheta\otimes
d\vartheta+\sin^2\vartheta d\varphi\otimes
d\varphi\right)\right\}$ and $T_{\rm RNK}=\frac{e^2 _{1,2}}{\kappa
{\bar r}^4}\left\{ dT\otimes dT-\right.$\linebreak $\left.{\cal
E}^2dR\otimes dR+{\bar r}^2\left(d\vartheta\otimes d\vartheta
+\sin^2\vartheta d\varphi\otimes d\varphi\right)\right\}$, thus
\be [T_{\mu\nu}]=0~~~~\Rightarrow~~~~~~ {k^2}=\frac{e^2
_{1,2}}{{\bar r}^4}. \label{A2} \ee Taking into account (\ref{A1})
and (\ref{A2}), we see that $\Lambda=\frac{{\bar r} ^2-e^2
_{1,2}}{{\bar r} ^4}$ and $e^2 _{1,2}=e^2=\frac{k^2}{(k^2+
\Lambda)^4}$, thus the charges observed from opposite entrances of
the wormhole coincide up to the sign.

The interior of such wormholes is a non-stationary region if
$k^2>\Lambda>-k^2$ or a static region if $\Lambda>k^2$. These
types of wormholes are observed in one universe as black holes, in
another universe (or on another spacetime sheet of the former
universe) as white holes, though there is no singularity which
should correspond to usual black holes, since they belong here to
NBR. Observers in these two adjacent RNK universes would conclude
that the wormhole has an electric charge with the same absolute
value, but opposite signs in different universes (or the similar
situation with the magnetic charge); they would also measure a
non-zero positive mass of the wormhole, but this mass in general
will be different for observers in different universes (together
with the different values of the cosmological constant
corresponding to the respective worlds).

It is comparatively easy to construct examples of Penrose
diagrams' hybridization resulting in a connection of two RNK
worlds {\it via} a NBR-wormhole. They show that the wormholes
under consideration are traversable only in one direction (one-way
wormholes) taking the traveller to another sheet of spacetime
(behind the future infinity of the abandoned world); this is also
visualized by diagrams using synchronous coordinates. Of course,
the Penrose diagrams' hybridization cannot be simply shown on one
piece of paper since the RNK singularities and adjacent sectors
require more space than there is at one's disposal on one sheet so
that one has to identify some boundaries of these sectors without
mixing them with those pertaining to the NBR-wormhole. Therefore
we do not show such hybridized diagrams here.

Naturally, the junction of RNK worlds {\it via} static part of
NBR-wormhole can be also done not on horizons, but in the outside
parts of RNK worlds and of NBR spacetime, thus permitting to
consider construction of two-way wormholes; in this case it is
natural to glue together only static regions of both spacetimes.
This requires the use of more complicated prescriptions for
junction, and we do not come in these details leaving them to
another publication. The NBR solution is chosen in this talk as a
convenient tool to construct wormholes since it already has the
necessary properties for modelling them due to the angular part of
the NBR metric.

\end{document}